\documentclass[a4paper,11pt]{article}
\usepackage{multirow} % Ensure this package is included

\pdfoutput=1 % if your are submitting a pdflatex (i.e. if you have
\usepackage{comment}
\usepackage{jcappub} % for details on the use of the package, please
\usepackage{hyperref}
\usepackage{fontawesome}

\title{Exponential quintessence with momentum coupling to dark matter}

\author[a,b]{Alkistis Pourtsidou} %and my mate ELM

\affiliation[a]{Institute for Astronomy, University of Edinburgh, \\ Royal Observatory, Blackford Hill, Edinburgh, EH9 3HJ, U.K.}
\affiliation[b]{Higgs Centre for Theoretical Physics, School of Physics and Astronomy, \\ Edinburgh EH9 3FD, UK}

\emailAdd{alkistis.pourtsidou@ed.ac.uk}

\abstract{ 
We present updated constraints on an interacting dark energy - dark matter model with pure momentum transfer, where dark energy is in the form of a quintessence scalar field with an exponential potential. We
run a suite of MCMC analyses using the DESI DR2 BAO measurements, in combination
with CMB data from \emph{Planck} and supernovae data from DESY5. In contrast to the standard case of uncoupled quintessence, we find that values for the potential's slope parameter $\lambda \geq \sqrt{2}$, which are conjectured by string theory scenarios, are not excluded. If $\lambda$ is fixed to such a value, we find that the data favour the negative coupling branch of the model, which is the branch exhibiting late-time growth suppression. 
We also derive 95\% upper limits on the sum of the neutrino masses, finding $\sum m_\nu < 0.06$ eV ($\sum m_\nu < 0.16$ eV) when $\lambda$ is fixed (varied). Our results motivate further studies on dynamical dark energy models that obey string theory bounds and can be constrained with cosmological observations.
}

\begin{document}
\maketitle
\flushbottom

% ------------------------------------------------------------- %
\section{Introduction}
% ------------------------------------------------------------- %

The standard cosmological model, $\Lambda$CDM, which postulates that dark energy is in the form of a cosmological constant $\Lambda$, is being challenged by DESI's Data Release 2 BAO measurements. Combined with CMB data from \emph{Planck} \cite{Planck:2019nip,Planck:2018lbu} and ACT \cite{ACT:2023kun}, and type Ia supernovae data from from Pantheon+ \cite{Scolnic:2021amr, Brout:2022vxf}, Union3 \cite{Rubin:2023ovl}, and DESY5 \cite{DES:2024jxu}, DESI's results point towards dynamical dark energy \cite{DESI:2025zgx,DESI:2025fii,DESI:2025wyn, DESI:2025gwf}. Assuming the Chevallier-Polarski-Linder parametrisation of the dark energy equation of state \cite{Chevallier:2000qy, Linder:2002et} 
\begin{equation}
  w(a) = w_0 + w_a(1-a) \, ,  
  \label{eq:CPL}
\end{equation}
the data combination with the strongest significance ($4.2\sigma$ preference over $\Lambda$CDM) is \cite{DESI:2025zgx}:

\begin{equation}
\left.\begin{array}{l}w_0=-0.752 \pm 0.057 \\ w_a=-0.86_{-0.20}^{+0.23}\end{array}\right\} \begin{aligned} & \text { DESI DR2 }+ \text { CMB } \\ & + \text { DESY5 } \, .
\label{eq:DESIbestfit}
\end{aligned}
\end{equation}
When also allowing the sum of the neutrino masses to vary, \cite{DESI:2025zgx} reported 95\% upper limits from the combination of DESI and CMB: $\sum m_\nu < 0.064$ eV assuming $\Lambda$CDM, which is close to the lower limit set by neutrino oscillations experiments \cite{Esteban:2024eli}, and $\sum m_\nu < 0.16$ eV for the $w_0w_a$ model. 
%$\sum m_\nu > 0.059$ eV.

Following the baseline results, additional studies were performed by the DESI collaboration, confirming this trend for both DESI DR1 and DR2 BAO measurements, but with the latter providing a stronger statistical evidence for dark energy dynamics \cite{DESI:2024mwx, DESI:2025wyn}. While the best-fit model implies rapidly evolving dark energy with a phantom crossing at $z \sim 0.4$, dark energy models like quintessence with $w(a)>-1$ at all times are not ruled out \cite{DESI:2025fii, DESI:2025wyn}. 

A key assumption of the suite of dark energy models tested by the DESI collaboration so far is that dark energy and dark matter are uncoupled. The motivation for this work is to extend these studies by testing a popular and phenomenologically interesting interacting dark energy model: coupled quintessence with pure momentum exchange, which we first introduced in \cite{Pourtsidou:2013nha}. Pure momentum transfer models are interesting for several reasons. As noted in \cite{Simpson:2010vh}, an `elastic scattering' type of interaction between dark matter and dark energy could be possible, considering the nonrelativistic velocities associated with dark matter and the low density of dark energy. Furthermore, this type of interaction is much less tightly constrained than most coupled dark energy models, which typically involve background energy exchange affecting the primary CMB (see e.g. \cite{Amendola:1999er, Xia_2009, Valiviita:2009nu, Pourtsidou:2013nha} and references therein). Models with pure momentum transfer are essentially unconstrained by the CMB and can also provide late-time growth suppression, which means they can be used to alleviate the (tentative) $S_8$ tension\footnote{This is the $\sim 2\sigma$ discrepancy between weak lensing measurements of the clustering amplitude, $S_8$, and the value inferred from the \emph{Planck} CMB measurements assuming the standard cosmological model, $\Lambda$CDM \cite{Heymans:2020gsg, KiDS:2020suj}. However, recent weak lensing re-analyses find $S_8$ values that are more consistent with \emph{Planck} \cite{Wright:2025xka}.} \cite{Pourtsidou:2016ico, Baldi:2016zom, DiValentino:2019ffd, Kase:2019veo, Vagnozzi:2023nrq,Tsedrik:2025cwc}. 

In this paper, we present cosmological constraints on such an interaction in light of the DESI DR2 BAO measurements  and their implications for dark energy and neutrinos. The paper is organised as follows: In \autoref{section:formalism} we review the coupled quintessence model under consideration. In \autoref{section:results} we describe the datasets and priors we use and then present the results of a suite of MCMC analyses using the DESI DR2 BAO data combined with CMB data from \emph{Planck} and supernovae data from DES. We conclude in \autoref{section:conclusions}.

% ------------------------------------------------------------- %
\section{Formalism}\label{section:formalism}
% ------------------------------------------------------------- %

In the formalism of \cite{Pourtsidou:2013nha, Pourtsidou:2016ico}, the quintessence - dark matter pure momentum transfer model under consideration is described by the Lagrangian:
\begin{equation}
\label{eq:LT3}
    L(n,Y,Z,\phi) = F(Y,Z,\phi) + f(n)\,.
\end{equation}
where $\phi$ is the quintessence scalar field, $n$ is the dark matter fluid number density, $Y=(1/2)\nabla_\mu \phi \nabla^\mu\phi$ is the kinetic term, and
\begin{equation}
    Z = u^\mu \nabla_\mu \phi 
\end{equation}
is a coupling between the gradient of the scalar field and the fluid velocity $u^\mu$.
We consider a coupled model of the form:
\begin{equation}
\label{eq:F}
    F = Y + V(\phi) + \beta Z^2\,,
\end{equation}
where $V(\phi)$ is the quintessence potential and $\beta Z^2$ is the coupling function, with $\beta$ the coupling parameter. 

The energy density and pressure of the scalar field are given by~\cite{Pourtsidou:2013nha, Pourtsidou:2016ico}
\begin{equation}
\label{eq:rhophi}
    \bar{\rho}_\phi = \left(\frac{1}{2}-\beta\right) \frac{\dot{\bar{\phi}}^2}{a^2} + V(\phi) \,,
\end{equation}
\begin{equation}
\label{eq:pphi}
    \bar{p}_\phi = \left(\frac{1}{2}-\beta\right) \frac{\dot{\bar{\phi}}^2}{a^2} - V(\phi) \,,
\end{equation}
where $\bar{\phi}$ is the background value of the scalar field and dots denote differentiation with respect to conformal time (note that the background part of $Z$ is given by $\bar{Z} = -\dot{\bar{\phi}}/a$). The scalar field obeys the background Klein-Gordon equation
\begin{equation}
\label{eq:sfeback}
    \ddot{\bar{\phi}}+2 \mathcal{H} \dot{\bar{\phi}}+\left(\frac{1}{1-2 \beta}\right) a^2 \frac{d V}{d \phi}=0 \, .
\end{equation}
The background energy density of the cold dark matter is not modified by this form of coupling:
\begin{equation}
    \dot{\bar{\rho}}_\text{c} + 3\mathcal{H}\bar{\rho}_\text{c} = 0\,.
\end{equation}
The cold dark matter density contrast $\delta_\text{c} = \delta\rho_\text{c}/\bar{\rho}_\text{c}$ also obeys the standard continuity equation:
\begin{equation}
\label{eq:delta0}
    \dot{\delta}_\text{c} = -k^2 \theta_\text{c} - \frac{1}{2}\dot{h}\,,
\end{equation}
while the velocity divergence $\theta_\text{c}$ obeys the modified Euler equation:
\begin{equation}
\label{eq:theta0}
    \dot{\theta}_c=-\mathcal{H} \theta_c+\frac{(6 \mathcal{H} \beta \bar{Z}+2 \beta \dot{\bar{Z}}) \varphi+2 \beta \bar{Z} \dot{\varphi}}{a\left(\bar{\rho}_c-2 \beta \bar{Z}^2\right)}\,,
\end{equation}
and the scalar field perturbation, $\varphi$, obeys
\begin{equation}
\begin{gathered}(1-2 \beta)(\ddot{\varphi}+2 \mathcal{H} \dot{\varphi})+\left(k^2+a^2 V_{\phi \phi}\right) \varphi \\ +\frac{1}{2} \dot{\bar{\phi}}(1-2 \beta) \dot{h}-2 \beta \dot{\bar{\phi}} k^2 \theta_c=0 \,.\end{gathered} 
\end{equation}
The perturbed Einstein field equations are not modified by the coupling and take their standard form.
For the quintessence potential, we choose the single exponential form
\begin{equation}
V(\phi) = V_0 e^{-\lambda \phi/M_{\rm pl}} \, ,
\label{eq:Vphi}
\end{equation}
with $M_{\rm pl}$ the reduced Planck mass.
Quintessence models have historically been the main candidates for dynamical dark energy \cite{Ratra:1987rm, Caldwell:1997ii, Carroll:1998zi, Copeland:2006wr}, and in particular exponential potentials are common in supergravity and string theory \cite{Barreiro:1999zs, Cicoli:2023opf, Andriot:2024jsh}. Observational constraints on quintessence models with a potential of the form of \autoref{eq:Vphi} can have important implications for string theory, as constructions where the theory is under perturbative control require $\lambda \geq \sqrt{2}$ \cite{Agrawal:2018own, Obied:2018sgi,Bedroya:2019snp, Rudelius:2021oaz}. 

The quintessence - dark matter momentum coupling model has been implemented in a publicly available version \href{https://github.com/Alkistis/class_IDE}{\faGithub} of the \texttt{CLASS} Boltzmann solver  \cite{Lesgourgues:2011re, Blas:2011rf}. The effect of the coupling in the CMB temperature and matter power spectra has been demonstrated in several papers (see e.g. Figures 1 and 2 in \cite{Pourtsidou:2016ico}). In summary, the effect of this type of coupling in the CMB temperature power spectrum is very small, only visible in the integrated Sachs-Wolfe effect on very large scales. The effects on the matter power spectrum are more significant: in general, for positive coupling ($\beta>0$) the growth is enhanced, while for the negative coupling case the growth is suppressed. These properties were used in \cite{Pourtsidou:2016ico} to demonstrate how pure momentum transfer in the dark sector can alleviate the $S_8$ tension. 

% ------------------------------------------------------------- %
\section{Cosmological constraints}\label{section:results}
% ------------------------------------------------------------- %

In the following, we explore the parameter space and constrain our model by performing a suite of MCMC analyses. To do this, we use our modified \texttt{CLASS} code and the \texttt{mcmc} sampler \cite{Lewis:2002ah, Lewis:2013hha} through their interface with \texttt{Cobaya} \cite{Torrado:2020dgo}. Our chains are converged when the Gelman-Rubin diagnostic \cite{1992StaSc...7..457G} $R-1 < 0.01$. The chains are analysed and plotted with \texttt{GetDist} \cite{Lewis:2019xzd}. We use the following datasets and likelihoods: \\
\textbf{CMB:} \emph{Planck} 2018 low-$\ell$ temperature and polarisation likelihood \cite{Aghanim:2019ame}, the \texttt{CamSpec} high-$\ell$ TTTEEE temperature and polarization likelihood using
\texttt{NPIPE} (\emph{Planck} PR4) data \cite{Rosenberg:2022sdy}, and the \emph{Planck} PR4 lensing likelihood \cite{Planck:2018lbu, Carron:2022eyg}. \\
\textbf{BAO:} Baryon Acoustic Oscillation (BAO) likelihood for all tracers from DESI DR2 \cite{DESI:2025zgx}. \\
\textbf{SN:} Likelihood for the DES-Y5 type Ia supernovae sample \cite{DES:2024jxu}. \\

We note that, when analysed independently assuming $\Lambda$CDM, the DESI DR2 BAO measurements and the DES Y5 SNe Ia data infer values of $\Omega_{\rm m}$ which are in moderate tension, at the level of $\simeq 2.9\sigma$ (see Figure 10 in \cite{DESI:2025zgx}). This tension is alleviated for dynamical dark energy. We have performed this check for our coupled quintessence model, finding good agreement between the recovered values of $\Omega_{\rm m}$ (well within $1\sigma$). Therefore, we can safely combine these datasets. We vary the standard cosmological parameters 
$$\{\omega_{\text {b}}, \omega_{\text {cdm }}, \theta_{\rm s}, A_{\rm s}, n_{\rm s}, \tau_{\text {reio }}\}$$ and the nuisance parameters required by the likelihoods we use. Here, $\omega_{\text {b}}$ is the physical baryon density, $\omega_{\text {cdm}}$ is the physical cold dark matter density, $\theta_{\rm s}$ is the angular scale of the sound horizon, $\tau_{\rm reio}$ is the optical depth of reionization, $A_{\rm s}$ is the amplitude, and $n_{\rm s}$ the tilt of the primordial power spectrum.
In all runs the initial conditions for the quintessence field are $\phi_i= 10^{-4}, \dot{\phi}_i=0$, but the cosmological evolution is insensitive to those \cite{Copeland:2006wr, Bhattacharya:2024hep}. The potential normalisation, $V_0$, is tuned by \texttt{CLASS} in order to match the dark energy density today and close the Friedmann equations, and we always assume a spatially flat Universe.  We will state the settings and priors for the quintessence potential parameter, $\lambda$, the coupling parameter, $\beta$, as well as the sum of the neutrino masses, $M_\nu \equiv \sum m_\nu/{\rm eV}$, for each case we consider in the subsections below.

\subsection{Coupled vs uncoupled quintessence}

We start with a comparison between the coupled model and uncoupled quintessence, with the latter corresponding to setting the coupling parameter $\beta=0$ in our modified \texttt{CLASS} code. For the coupled model, the range of $\beta$ values we consider is important. There is a theoretical prior which does not allow us to consider the case $\beta \geq 1/2$, due to ghost pathologies in this branch of the model \cite{Pourtsidou:2013nha}. We are free to take any value $\beta < 1/2$, but the fact that it is a dimensionless parameter in the model's Lagrangian suggests that its magnitude should be $\mathcal{O}(1)$. Following this reasoning, we choose a prior range for our coupling parameter such that $-2.0 \leq \beta < 0.5$. 
For the potential's slope parameter we consider a flat prior $0 \leq \lambda \leq 2.1$ as in \cite{Pourtsidou:2016ico}. The total neutrino mass is kept fixed, $M_\nu=0.06$. 

\begin{figure}[h]
    \centering
    \includegraphics[width=0.6\columnwidth]{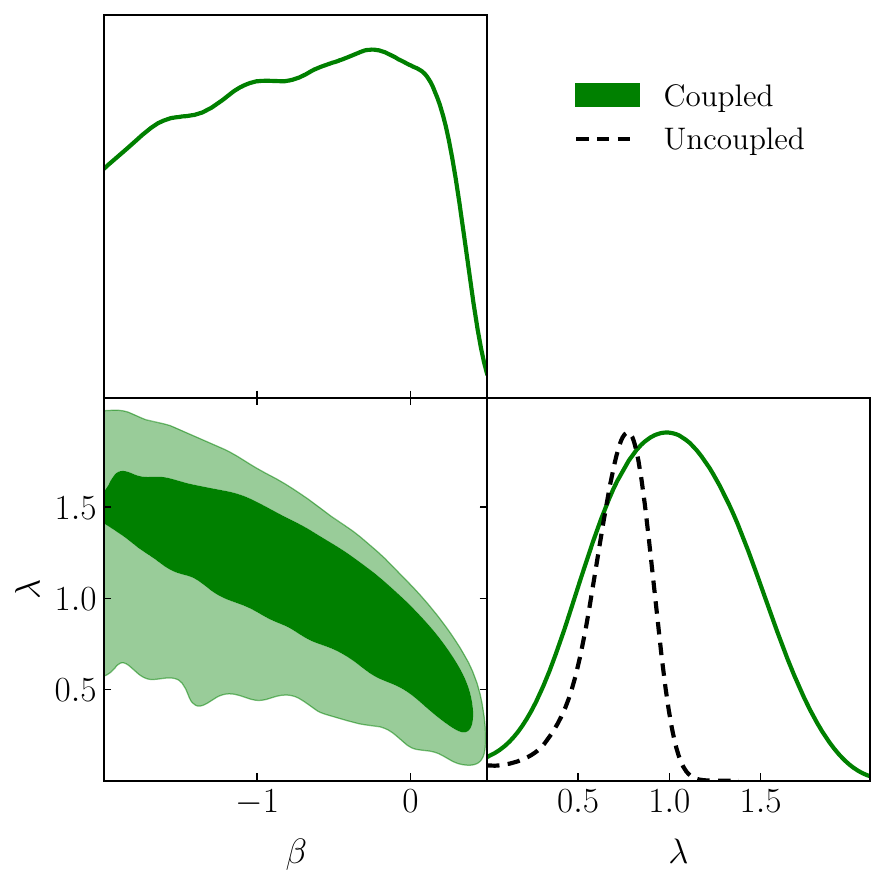}
    \caption{One dimensional posterior distributions of the parameters $\{\beta,\lambda\}$ together with the contours containing 68\% and 95\% of the posterior probability for the momentum coupling model (solid green lines) vs uncoupled quintessence (dashed black line), where the latter corresponds to fixing the coupling parameter $\beta=0$. We notice that the coupled case allows $\lambda \geq \sqrt{2}$, which is conjectured by string theory scenarios.}
    \label{fig:coupled-uncoupled}
\end{figure}

Our results are shown in \autoref{fig:coupled-uncoupled} and \autoref{tab:coupled-uncoupled}. The uncoupled exponential quintessence case has also been constrained with DESI BAO data in recent works \cite{Bhattacharya:2024hep, Ramadan:2024kmn, Akrami:2025zlb}, and our results are in agreement with them\footnote{For constraints with pre-DESI data and their implications for dark energy in string theory, see \cite{Akrami:2018ylq, Raveri:2018ddi, Heisenberg:2018yae} and references therein.}. We see that the data prefer a nonzero $\lambda$ in both the coupled and uncoupled quintessence cases. We notice that the uncoupled case excludes `string theory motivated' values $\lambda \geq \sqrt{2}$, but the coupled case allows for them. At first sight, this might be considered expected since we have opened up the parameter space; however, this has not been the case for another popular extension of the exponential quintessence model allowing for nonzero spatial curvature \cite{Bhattacharya:2024hep}. In our model, this feature is due to the explicit $\lambda - \beta$ degeneracy, and the fact that the coupling parameter $\beta$ is essentially unconstrained. For completeness, we have included the full cosmological parameter contours for these cases in \autoref{sec:appendix} (\autoref{fig:full-coupled-uncoupled}). We have also run the \texttt{BOBYQA} minimiser \citep{cartis2018improvingflexibilityrobustnessmodelbased,Cartis_2021, Powell2009BOBYQA} to evaluate the goodness of fit, calculating the $\chi^2$ starting from the maximum a posteriori (MAP) points of each of the chains in the MCMC sampling. We considered the coupled and uncoupled quintessence models, as well as $w_0w_a$ and $\Lambda$CDM. For $\Lambda$CDM, we have $\chi^2 \simeq 12639$. Computing the $\Delta \chi^2$ relative to $\Lambda$CDM we find $\Delta \chi^2 \simeq -20$, $-12$, and $-12$, for $w_0w_a$, coupled, and uncoupled quintessence, respectively. 
This shows that all dynamical dark energy models give better fits than $\Lambda$CDM, with $w_0w_a$ performing best.

\begin{figure}[h]
    \centering
    \includegraphics[width=0.7\columnwidth]{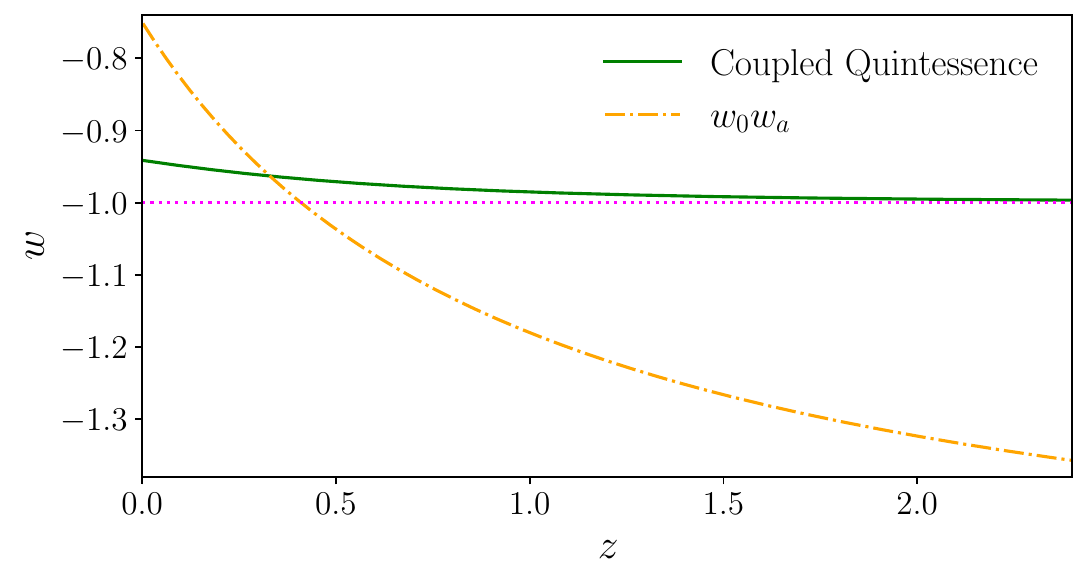}
    \caption{Dark energy equation of state for the coupled quintessence model with $(\beta,\lambda)=(-0.8,1)$ from \autoref{tab:coupled-uncoupled}, compared to the 
    CPL parametrisation with $(w_0,w_a)=(-0.75,-0.86)$ from \autoref{eq:DESIbestfit}.}
    \label{fig:wDE}
\end{figure}

It is also useful to note that the dark energy equation of state for the coupled quintessence model is given by \cite{Pourtsidou:2016ico}
\begin{equation}
w \equiv \frac{p_\phi}{\rho_\phi} = \frac{\left(\frac{1}{2}-\beta\right) \frac{\dot{\bar{\phi}}^2}{a^2}-V(\phi)}{\left(\frac{1}{2}-\beta\right) \frac{\dot{\bar{\phi}}^2}{a^2}+V(\phi)} \, ,
\end{equation}
with $\beta < 1/2$ to avoid ghost and strong coupling pathologies. This means that $w$ cannot be in the phantom regime (see \autoref{fig:wDE}). For more details on the dependence of the $w$ evolution on the coupling parameter $\beta$, we refer the reader to \citep{Pourtsidou:2016ico, Chamings:2019kcl}. \\
\begin{table}[h]
\centering
\begin{tabular}{|c |c | c |c |} 
 \hline
\textbf{Fixed} $M_\nu=0.06$  & \textbf{Coupled} & \textbf{Coupled, fixed $\lambda$} &\textbf{Uncoupled} \\ [0.5ex] 
 $100~\omega_{\rm b}$ & $2.23 \pm 0.01$ & $2.23 \pm 0.01$ & $2.23 \pm 0.01$ \\ 
 $\omega_\text{cdm}$ & $0.117\pm 0.001$ & $0.117\pm 0.001$ & $0.117\pm 0.001$ \\
 $10^4\theta_{\rm s}$ & $104.20\pm 0.02$ & $104.20\pm 0.02$ & $104.20\pm 0.02$
 \\
 $n_\text{s}$ & $0.969\pm 0.004$ & $0.969\pm 0.003$  & $0.969\pm 0.003$ \\
 $\tau_\text{reio}$ & $0.06\pm 0.01$ & $0.06\pm 0.01$ & $0.06\pm 0.01$ \\
 $\sigma_{8}$ & $0.777_{-0.015}^{+0.026}$ & $0.752\pm 0.006$ & $0.793\pm 0.008$ \\
 $H_0$ & $66.9\pm 0.6$ & $66.6_{-0.3}^{+0.4}$  & $66.9\pm 0.6$ \\
 $\lambda$ & $1.0\pm 0.4$ & 1.5 & $0.7_{-0.1}^{+0.2}$ \\
 $\beta$ & $-0.8_{-0.6}^{+1.0}$ & $-1.4_{-0.5}^{+0.3}$ & 0 \\ [1ex]
% $M_\nu$ & 0.06 & 0.06 & 0.06 \\ [1ex] 
 \hline
\end{tabular}
\caption{\label{tab:coupled-uncoupled} Cosmological parameters for the coupled and uncoupled quintessence models considered in this work, with fixed $M_\nu=0.06$. We quote upper and lower values at the 68\% confidence level.}
\end{table}

\begin{figure}[h]
    \centering
    \includegraphics[width=0.6\columnwidth]{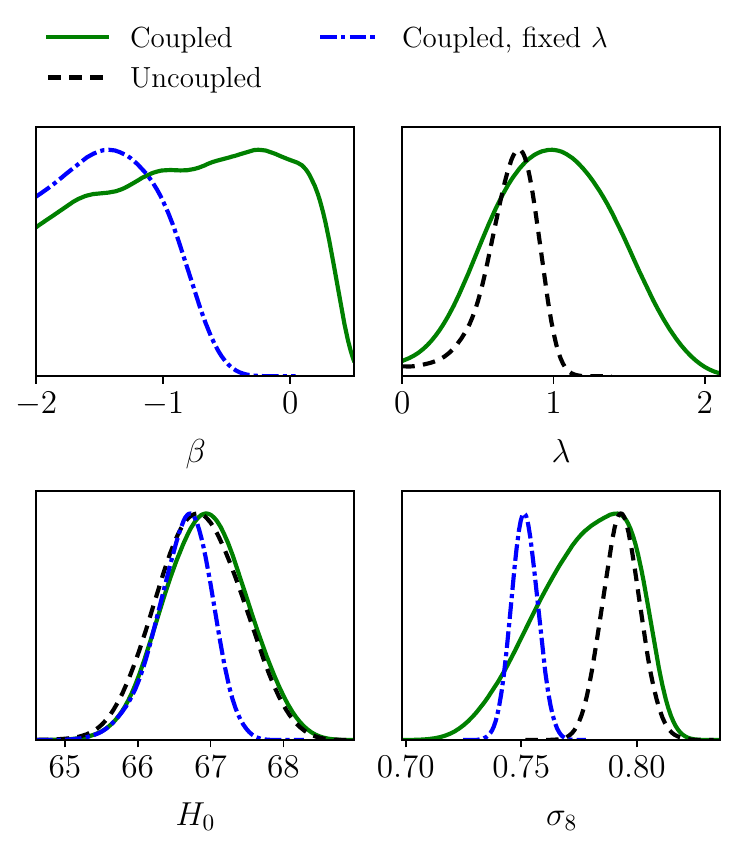}
    \caption{One dimensional posterior distributions of the parameters $\{\beta,\lambda\}$, and the derived parameters $\{H_0, \sigma_8\}$, for the momentum coupling model with varying (solid green lines) vs fixed $\lambda=1.5$ (dashed-dotted blue lines), and uncoupled quintessence (dashed black lines). Note that in all cases shown here the total neutrino mass is fixed, $M_\nu=0.06$.
    }
    \label{fig:coupled_vs_coupled_fixed_lambda}
\end{figure}

Given these findings, we will now fix the quintessence potential parameter to a `string theory motivated' value $\lambda=1.5$ and repeat the analysis keeping all other parameters and priors the same. Our results are shown in \autoref{fig:coupled_vs_coupled_fixed_lambda} and \autoref{tab:coupled-uncoupled}. Interestingly, in the fixed $\lambda=1.5$ case the data exclude the positive coupling branch of the model and prefer the negative branch, $\beta<0$. As demonstrated in \cite{Pourtsidou:2016ico} and \cite{Chamings:2019kcl}, the negative coupling branch is exhibiting late-time growth suppression (in contrast to the positive coupling branch, which leads to growth increase), and is able to alleviate the $S_8$ tension. Indeed, as we see in \autoref{fig:coupled_vs_coupled_fixed_lambda}, the case with fixed $\lambda=1.5$ prefers negative $\beta$ values and predicts a low $\sigma_8$.

\subsection{Coupled quintessence with varying neutrino mass}

\begin{figure}[h]
    \centering
    \includegraphics[width=0.9\columnwidth]{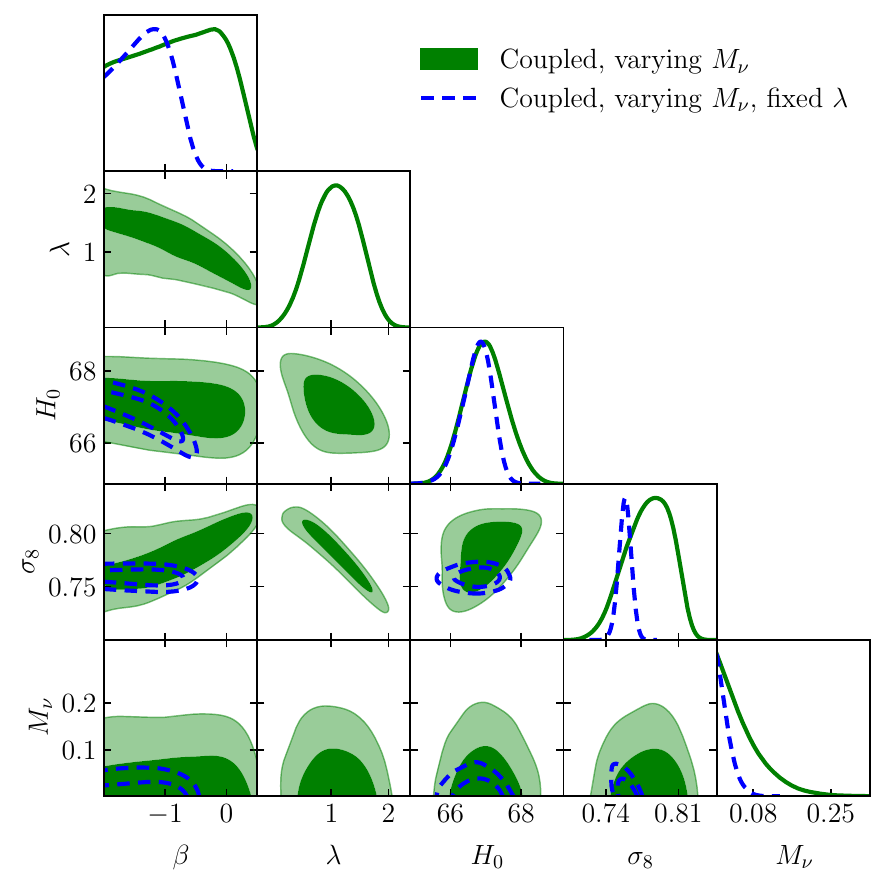}
    \caption{One dimensional posterior distributions of the parameters $\{\beta, \lambda, H_0,\sigma_8, M_\nu\}$ together with the contours containing 68\% and 95\% of the posterior probability for the momentum coupling model when allowing the total neutrino mass to vary. We see that allowing $\lambda$ to vary (solid green lines) significantly relaxes the neutrino bounds compared to the case where $\lambda=1.5$ (dashed blue lines).}
    \label{fig:coupled-Mnu}
\end{figure}

We will now also allow the total neutrino mass, $M_\nu$, to vary, and impose the physical prior $M_\nu>0$. We start from the coupled quintessence case of the previous subsection, fixing $\lambda=1.5$. The results are shown in \autoref{fig:coupled-Mnu} and \autoref{tab:coupled-Mnu}. Similarly to the baseline DESI results \cite{DESI:2025zgx, DESI:2025ejh}, the $M_\nu$ posterior probability would peak at negative neutrino masses if not restricted by the positive mass prior. To explore the $M_\nu$ behaviour further, we also consider the case where $\beta$, $\lambda$, and $M_\nu$ are allowed to vary. As we see in \autoref{fig:coupled-Mnu}, allowing $\lambda$ to vary significantly relaxes the neutrino bounds compared to the previous, fixed $\lambda$ case. The 95\% upper bound is $M_\nu < 0.06$ for the fixed $\lambda$ case, and $M_\nu < 0.16$ for the varied $\lambda$ case. 

\begin{table}[h]
\centering
\begin{tabular}{|c |c | c |} 
 \hline
 & \textbf{Coupled} & \textbf{Coupled, fixed $\lambda$} \\ [0.5ex] 
 $100~\omega_{\rm b}$ & $2.23 \pm 0.01$ & $2.23 \pm 0.01$ \\ 
 $\omega_\text{cdm}$ & $0.118\pm 0.001$ & $0.118\pm 0.001$ \\
 $10^4\theta_{\rm s}$ & $104.20\pm 0.02$ & $104.20\pm 0.02$ 
 \\
 $n_\text{s}$ & $0.968\pm 0.004$ & $0.969\pm 0.004$  \\
 $\tau_\text{reio}$ & $0.06\pm 0.01$ & $0.06\pm 0.01$ \\
 $\sigma_{8}$ & $0.781^{+0.026}_{-0.019}$ & $0.758\pm 0.006$ \\
 $H_0$ & $67.0\pm 0.6$ & $66.7^{+0.5}_{-0.4}$ \\
 $\lambda$ & $1.1\pm 0.4$ & $1.5$ \\
 $\beta$ & $-0.8^{+0.9}_{-0.6}$ & $-1.3\pm 0.4$ \\ 
 $M_\nu$ & $< 0.16$ & $< 0.06$ \\ [1ex] 
 \hline
\end{tabular}
\caption{\label{tab:coupled-Mnu} Cosmological parameters for the coupled quintessence models considered in this work, where the total neutrino mass $M_\nu$ is also allowed to vary. We quote upper and lower values at the 68\% confidence level, except for $M_\nu$ where we quote the 95\% upper bound for comparability with previous works. 
The lower limit set by neutrino oscillations experiments is $M_\nu > 0.059$.
}
\end{table}

\begin{figure}[h]
    \centering
    \includegraphics[width=0.8\columnwidth]{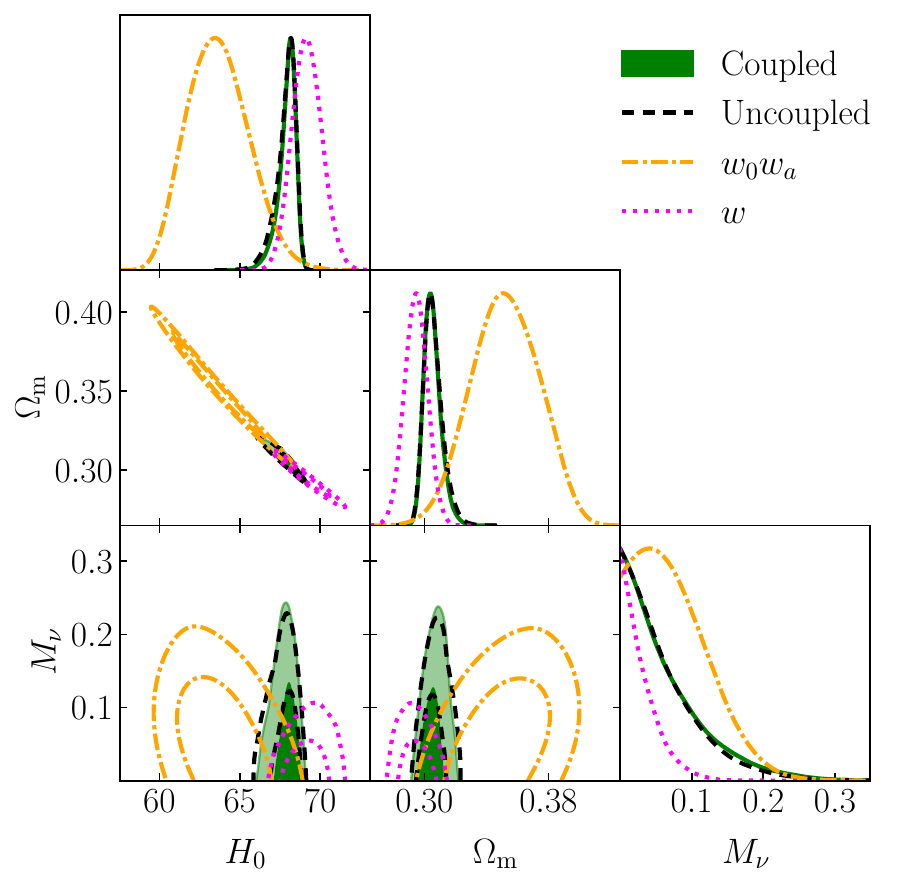}
    \caption{One dimensional posterior distributions of the parameters $\{H_0,\Omega_{\rm m}, M_\nu\}$ together with the contours containing 68\% and 95\% of the posterior probability for the momentum coupling model (solid green lines) and uncoupled quintessence (dashed black lines), as well as the $w_0w_a$ and constant $w$ parametrisations (dotted dashed orange lines and dotted magenta lines, respectively). Neither quintessence models are able to reproduce the positive peak in the $M_\nu$ posterior of the $w_0w_a$ parametrisation, but they allow for much more relaxed $M_\nu$ bounds than $w$CDM. Note that for the MCMC runs shown here we have used the CMB and BAO datasets only (without SN).}
    \label{fig:Mnu-w0wa}
\end{figure}

Given the baseline DESI results \cite{DESI:2025zgx}, it is also worth comparing our model's constraints with the $w_0w_a$ parametrisation, which allows phantom crossing. For this comparison, we remove the DESY5 supernovae dataset from our MCMC runs; in the $w_0w_a$ case this results in the
peak of the 1D marginalized posterior to be recovered in the
positive mass range \cite{DESI:2025zgx}. The results are shown in \autoref{fig:Mnu-w0wa} for the $\{H_0,\Omega_{\rm m},M_\nu\}$ parameters, with $\Omega_{\rm m}$ the total matter density. We see that the bounds on $M_\nu$ are very similar for the coupled and uncoupled cases, which means that the potential parameter $\lambda$ has a more significant effect than the coupling parameter $\beta$.
However, neither of the quintessence models is able to reproduce the positive peak of the $M_\nu$ posterior of the much more flexible $w_0w_a$ parametrisation, whose samples lie mostly within the `Quintom B' regime, with $w<-1$ in the past and $w>-1$ today \cite{Feng:2004ad, DESI:2025wyn}. We also see that in the case of coupled and uncoupled quintessence, DESI prefers larger values of $H_0$ and smaller values of $\Omega_{\rm m}$ compared to the $w_0w_a$ case. This is analogous to the trends seen in the DESI analysis of flat $\Lambda$CDM and $w$CDM compared to $w_0w_a$CDM \cite{DESI:2025zgx}. To demonstrate this, in \autoref{fig:Mnu-w0wa} we also plot the results for the corresponding $w$CDM analysis. We see that both quintessence models allow for relaxed neutrino mass bounds compared to the constant $w$ case.

%

% ------------------------------------------------------------- %
\section{Conclusions}\label{section:conclusions}
% ------------------------------------------------------------- %

We presented updated cosmological constraints on exponential quintessence with momentum coupling to dark matter, and compared the results with the standard (uncoupled) exponential quintessence model. Our main motivation was the observational evidence for evolving dark energy when combining the latest DESI DR2 BAO measurements with CMB data from Planck and supernovae data. 

Studying exponential quintessence models of dark energy also has theoretical motivation. For example, the string theory swampland conjectures suggest that effective field theories with exact (metastable) de Sitter vacua cannot be UV-completed, which, taken at face value, rules out $\Lambda$CDM.
In this work, we first confirmed previous findings on the allowed values of the potential slope for exponential quintessence, which are in tension with the  trans-Planckian
censorship conjecture and the strong de Sitter conjecture \cite{Bedroya:2019snp, Rudelius:2021oaz}. We then found that postulating a pure momentum coupling between quintessence and dark matter allows for $\lambda \geq \sqrt{2}$, alleviating this tension. Looking ahead, this result motivates further studies on evolving dark energy and modified gravity models that can relieve this tension and satisfy observational constraints (see, for example, \cite{Heisenberg:2019qxz,Brahma:2019kch, vandeBruck:2019vzd, Agrawal:2019dlm, Bedroya:2025fwh, Shlivko:2024llw, Giare:2024smz, Andriot:2025los, Anchordoqui:2025fgz, Wolf:2025jed, Bayat:2025xfr}).

Fixing $\lambda=1.5$ we found that the data favour the negative coupling branch of our model, $\beta < 0$, which is known for being able to resolve the $S_8$ tension. We also derived 95\% upper limits on the sum of the neutrino masses, finding $\sum m_\nu < 0.06$ eV when $\lambda=1.5$ and $\sum m_\nu < 0.16$ eV  when $\lambda$ is allowed to vary. Considering the CMB and BAO data combination only, we found that neither uncoupled nor coupled quintessence is able to reproduce the positive neutrino mass peak in the posterior found for the $w_0w_a$ parametrisation. 

In future work, it will be interesting to constrain the Dark Scattering model of \cite{Simpson:2010vh}, which can have a $w_0w_a$ background allowing for quintom/phantom evolution, and also includes pure momentum exchange. For both Dark Scattering and our coupled quintessence model, it is crucial to have accurate nonlinear modelling prescriptions in order to exploit the full constraining power of Stage IV surveys like DESI, Euclid \cite{Euclid:2024yrr}, and Rubin-LSST \cite{LSSTDarkEnergyScience:2018jkl}. This can be achieved using bespoke N-body simulations \cite{Baldi:2014ica, Palma:2023ggq, Luo:2025szq} and semi-analytic approaches like the halo model reaction \cite{Cataneo:2018cic, Bose:2021mkz, Carrilho:2021rqo}. 

\acknowledgments

AP's research is supported by a UK Research and Innovation Future Leaders Fellowship [grant MR/X005399/1]. We are grateful to Suddhasattwa Brahma, Pedro Carrilho and Maria Tsedrik for their help and feedback. We acknowledge use of the Cuillin computing cluster, Royal Observatory, University of Edinburgh. For the purpose of open access, the author has applied a Creative Commons Attribution (CC BY) licence to any Author Accepted Manuscript version arising from this submission. 

\bibliographystyle{JHEP}
\bibliography{mybib.bib}

\clearpage

\appendix
\section{Full posterior distributions}
\label{sec:appendix}

\begin{figure}[h]
    \centering
    \includegraphics[width=0.9\columnwidth]{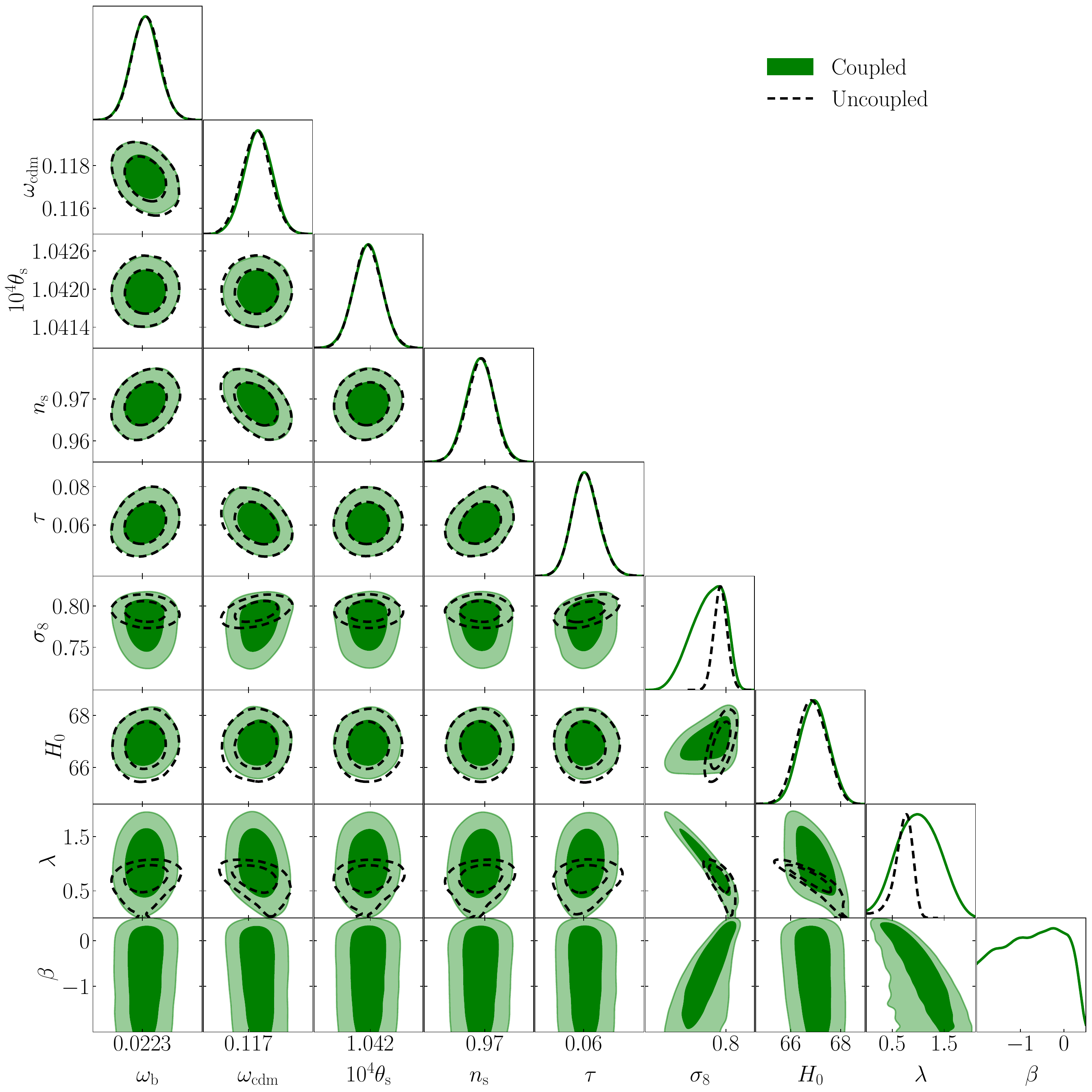}
    \caption{One dimensional posterior distributions of the cosmological parameters together with the contours containing 68\% and 95\% of the posterior probability for the momentum coupling model (solid green lines) vs uncoupled quintessence (dashed black lines) for the cases of \autoref{fig:coupled-uncoupled}.}
    \label{fig:full-coupled-uncoupled}
\end{figure}

\end{document}